\documentclass[preprint,11pt]{elsarticle}

\usepackage{amssymb}
\usepackage{adjustbox}
\usepackage{pdflscape}
\usepackage{natbib}
\usepackage{amsmath}
\usepackage{bm} 
\usepackage{multirow}
\usepackage{graphics}
\usepackage{epsfig}
\usepackage{pdfpages}
\usepackage{statmath}
\usepackage{amssymb}
\usepackage{algpseudocode}
\usepackage{algorithm}
\algnewcommand\algorithmicforeach{\textbf{for each}}
\algdef{S}[FOR]{ForEach}[1]{\algorithmicforeach\ #1\ \algorithmicdo}
\algnewcommand{\Inputs}[1]{%
  \State \textbf{Inputs:}
  \Statex \hspace*{\algorithmicindent}\parbox[t]{.8\linewidth}{\raggedright #1}
}
\algnewcommand{\Initialize}[1]{%
  \State \textbf{Initialize:}
  \Statex \hspace*{\algorithmicindent}\parbox[t]{.8\linewidth}{\raggedright #1}
}

\journal{}
\begin{document}
\begin{frontmatter}


\title{Joint Self-Supervised and Supervised Contrastive Learning for Multimodal MRI Data: Towards Predicting Abnormal Neurodevelopment}

\author[1,4]{Zhiyuan Li}
\author[1,2,3,5]{Hailong Li}
\author[4]{Anca L. Ralescu}
\author[1,2,5]{Jonathan R. Dillman}
\author[7]{Mekibib Altaye}
\author[1,5,6]{Kim M. Cecil}
\author[3,6]{Nehal A. Parikh}
\author[1,2,3,5]{Lili He \corref{cor1}}
\cortext[cor1]{Corresponding author: lili.he@cchmc.org (Lili He)}

\address[1]{Imaging Research Center, Department of Radiology, Cincinnati Children’s Hospital Medical Center, Cincinnati, OH, USA}
\address[2]{Artificial Intelligence Imaging Research Center, Cincinnati Children’s Hospital Medical Center, Cincinnati, OH, USA}
\address[3]{Neurodevelopmental Disorders Prevention Center, Perinatal Institute, Cincinnati Children's Hospital Medical Center, Cincinnati, OH, USA}
\address[4]{Department of Computer Science, University of Cincinnati, Cincinnati, OH, USA}
\address[5]{Department of Radiology, University of Cincinnati College of Medicine, Cincinnati, OH, USA}
\address[6]{Department of Pediatrics, University of Cincinnati College of Medicine, Cincinnati, OH, USA}
\address[7]{Biostatistics and Epidemiology, Cincinnati Children’s Hospital Medical Center, Cincinnati, OH, United States}

\begin{abstract}
The integration of different imaging modalities, such as structural, diffusion tensor, and functional magnetic resonance imaging, with deep learning models has yielded promising outcomes in discerning phenotypic characteristics and enhancing disease diagnosis. The development of such a technique hinges on the efficient fusion of heterogeneous multimodal features, which initially reside within distinct representation spaces. Naively fusing the multimodal features does not adequately capture the complementary information and could even produce redundancy. In this work, we present a novel joint self-supervised and supervised contrastive learning method to learn the robust latent feature representation from multimodal MRI data, allowing the projection of heterogeneous features into a shared common space, and thereby amalgamating both complementary and analogous information across various modalities and among similar subjects. We performed a comparative analysis between our proposed method and alternative deep multimodal learning approaches. Through extensive experiments on two independent datasets, the results demonstrated that our method is significantly superior to several other deep multimodal learning methods in predicting abnormal neurodevelopment. Our method has the capability to facilitate computer-aided diagnosis within clinical practice, harnessing the power of multimodal data.
\end{abstract}

\end{frontmatter}

\section{Introduction}
Neurodevelopmental abnormalities pose a significant risk to infants born very prematurely ($<32$ weeks gestational age). However, diagnosing or predicting deficits before the age of 3 years remains challenging. Accurate early prediction models are urgently needed to facilitate risk stratification and enable timely interventions, thus maximizing the well-being of children and their families. Advances in magnetic resonance imaging (MRI) and deep learning provide means to address this unmet need.

Different MRI modalities, such as structural MRI (sMRI), diffusion tensor imaging (DTI), and functional MRI (fMRI) can provide complementary information about the anatomy, neural pathways, and functions of the brain \cite{kidwell2003beyond}. sMRI studies brain static anatomical properties utilizing the magnetic properties of protons in water molecules \cite{frisoni2010clinical}. DTI maps the white matter pathways in the brain to reveal the microstructural organization of white matter tracts and their integrity by measuring the diffusion of water molecules \cite{jones2010diffusion}. fMRI detects changes in blood flow and oxygenation that occur in response to neural activity. It depicts the functional organization of the brain and shows how different brain regions are functionally connected to one another  \cite{friston1994analysis}. Research has shown that integrating multimodal information is more effective than using a single modality for identifying phenotypic characteristics and improving the prediction/diagnosis of neurological and neurodevelopmental impairments in very preterm infants \cite{dai2020multimodal,lee2022multimodal}.

Multimodal learning aims to construct artificial intelligence (AI) models that can analyze and integrate relevant features extracted from diverse data modalities, with the goal of performing various tasks such as classification and regression \cite{ramachandram2017deep}. Multimodal learning has seen significant advancements in recent years, with most prior studies being motivated by one primary driver: the complementary information provided by each modality. This allows the AI models to leverage the unique strengths of each modality and gain a more comprehensive understanding of the input data/features \cite{wang2022deep}. Conventionally, a wide range of kernel-based machine learning algorithms has been proposed to summarize and fuse complementary information through linear and nonlinear combination methods \cite{poria2015deep,wen2017multi}. With the advancement of deep learning, deep multimodal fusion methods \cite{wang2020deep,huang2020multimodal,he2021deep,boulahia2021early} have become increasingly popular. These methods extract latent feature representations/embeddings for each modality using deep neural networks and then combine them in different ways such as concatenation \cite{radu2018multimodal,he2021deep}, canonical correlation-based analysis (CCA) \cite{liu2019multimodal,yuan2019joint,puyol2022multimodal}, and attention \cite{he2023co,jha2023gaf}. However, these heterogeneous multimodal feature representations are originally located in different representation spaces, and naively fusing them does not appropriately capture the complementary information and could even produce redundancy information \cite{bakkali2023vlcdoc}. Self-supervised contrastive learning techniques have been proposed to address this issue by mapping heterogeneous feature representations into a common representation space, where they can be more effectively combined. These methods aim to identify similarities and differences between different modalities and leverage this information to create more informative and robust representations. For example, modality-invariant methods \cite{li2020self,sun2023modality}  aim to learn representations that are invariant to modality-specific factors, while CLIP-based methods \cite{radford2021learning,sanghi2022clip,wang2022clip} leverages contrastive learning to create a shared representation space for images and text. Other methods such as ContIG \cite{taleb2022contig}, ConVIRT \cite{zhang2022contrastive}, and VATT \cite{akbari2021vatt} use variants of attention mechanisms to more effectively combine information from different modalities.  

In the field of classification, it has been acknowledged that mining shared information across subjects from the same class is the essence of enhancing the performance of classification models \cite{huang2019multimodal,zhang2021supervised,zhu2022multimodal}. Supervised contrastive learning \cite{khosla2020supervised} has merged as a powerful representation learning technique, which enhances classification performance by emphasizing both the similarities and differences between subjects. By mapping similar subjects (with the same class labels) close together and dissimilar subjects (with different class labels) far apart in a common space, this technique can create latent space feature representations that are particularly effective for downstream classification tasks \cite{hoffer2015deep}. In medicine, this strategy has been widely applied, including through the use of Siamese-based \cite{aderghal2017classification,rossi2020multi}, Triplet-based \cite{yu2021multimodal,zhu2022multimodal}, and SupCon-based \cite{zhang2023multi} methods.

By leveraging the strengths of the abovementioned different contrastive fusion methods, we propose a novel joint self-supervised and supervised contrastive learning method. Our method aims to learn an enhanced multimodal feature representation by amalgamating both complementary information among different modalities via \emph{cross-modality-complementary (CMC) features learning} and shared information among similar subjects via \emph{cross-subject-similarity (CSS) features learning}. \emph{CMC features learning} brings together the multimodal feature representations of the same individual and pushes apart the multimodal feature representations of different individuals in the feature representation space. This helps our model to reduce redundant feature learning and enhance the complementary semantics among different modalities. In addition, our \emph{CSS features learning} enhances the alignment of similar subjects by minimizing the distance between their multimodal feature representations maximizing the distances among subjects from the same class, and maximizing the distances among subjects from different classes. This helps our model to identify commonalities among subjects and generalize to new subjects. The proposed method has the potential to improve the performance of neurodevelopment prediction tasks by leveraging complementary and shared information in multimodal MRI data. To demonstrate the effectiveness of our method, we implemented our method for the early prediction of abnormal neurodevelopmental outcomes in very preterm infants using two independent datasets. Our study makes the following contributions:
\begin{enumerate}
\item	We propose a novel joint self-supervised and supervised contrastive learning method that effectively captures complementary information and enhances the synergistic effect created across modalities and subjects. 

\item	Our learning objective loss combines cross-modality-complementary (CMC) and CSS (CSS) loss functions. By optimizing CMC loss, our method brings the multimodal features of the same subject closer and those of different subjects mutually exclusive, reducing redundancy and enhancing the complementary semantics among different modalities. By optimizing CSS loss, our method pulls the multimodal features of subjects from the same class closer and pushes away those of subjects from different classes, thus enhancing the alignment of similar subjects and generalizing them to new subjects. 

\item  Our extensive experiments demonstrate the superiority of our proposed method over other state-of-the-art deep multimodal learning, self-supervised, or supervised contrastive learning approaches. 
\end{enumerate}

\section{Related Work}
In this section, we provide a review of related literature on multimodal fusion methods and multimodal contrastive learning with a focus on their application to medical imaging.

\subsection{Multimodal Fusion Model}
Multimodal fusion models have been extensively studied for various medical imaging-related tasks to combine complementary information extracted from different modalities. The most common approach in existing works is to map input from different modalities to their corresponding feature representation spaces and aggregate them as a high-level fused feature representation. As discussed in the introduction, notable multimodal fusion methods can be categorized into concatenation, canonical correlation-based analysis (CCA), and attention. Using multimodal feature concatenation, He et al. \cite{he2021deep} proposed an end-to-end deep multimodal model that fused T2-weighted anatomical MRI, DTI, resting state fMRI (rs-fMRI), and clinical data to predict neurodevelopmental deficits. Tang et al. \cite{tang2020deep} used the concatenation of multimodal features from fMRI image volume and its extracted ROI time series to predict autism disorder. Joo et al. \cite{joo2021multimodal} concatenated high dimensional features from clinical information, T1- and T2-weighted MRI for the prediction of pathological complete response to neoadjuvant chemotherapy in breast cancer. 

CCA approach \cite{yang2019survey} uses product operation, which maximizes the correlation between two sets of variables, to capture the common information across multiple modalities. For instance, Lei et al. \cite{gao2017discriminative} fused MRI and positron emission tomography (PET) features by CCA and developed a discriminative learning model for Alzheimer’s disease prediction. Similarly, Subramanian et al. \cite{subramanian2021multimodal} proposed a multimodal fusion method that projects gene expressions and histology data to well-correlated spaces using CCA for breast cancer survival prediction. Puyol-Anton et al. \cite{puyol2022multimodal} applied the CCA strategy in a multimodal learning framework to learn the relationship between 2D cardiac magnetic resonance and 2D echocardiography data for predicting cardiac resynchronization therapy response. 

Attention-based multimodal learning methods consider the high-order information extracted from multimodal features and explore the latent correlation among the attention maps. For example, Song et al. \cite{song2022cross} developed a cross-attention multimodal method for correlating transrectal ultrasound features and MRI features in an image registration task. Dalmaz et al. \cite{dalmaz2022resvit} proposed a ResViT that employed an aggregated residual self-attention transformer to integrate multimodal MRI and CT images for medical image synthesis tasks.

\subsection{Self-Supervised Contrastive Learning for Multimodal Data}
In recent years, contrastive learning has emerged as a dominant approach in the representation learning area. Various advanced methods, such as Invariant\cite{ye2019unsupervised}, Moco v1-v3 \cite{he2020momentum,chen2020improved}, SimCLR \cite{chen2020simple}, BOYL \cite{grill2020bootstrap}, and SimSiam \cite{chen2021exploring}, have achieved superior performance in the medical imaging domain \cite{li2020self,liang2020computer,fedorov2021self}. These contrastive learning methods have been applied to multimodal data to map the heterogeneous features from different modalities into a common space for capturing complementary information and reducing redundancy. In particular, Li et al. \cite{li2020self} proposed a self-supervised modality-invariant method for retinal disease diagnosis by incorporating color fundus images, corresponding transformed color fundus images, and fundus fluorescein angiography together. Zhang et al. \cite{zhang2022contrastive} learned a hybrid representation of paired X-rays and their corresponding medical notes for pneumonia detection by maximizing the agreement between feature representations of images and text pairs. Taleb et al. \cite{taleb2022contig} introduced a self-supervised contrastive learning method, ContIG, by aligning feature representations of medical images and various genetic data for cardiovascular risk prediction and diabetic retinopathy detection. Zhang et al. \cite{zhang2023multi} developed a semi-supervised contrastive mutual learning (Semi-CML) and a soft pseudo-label re-learning (PReL) method to bridge the semantic gaps among different brain imaging modalities (CT, PET, and sMRI) for medical image segmentation. Fischer et al. \cite{fischer2023self} combined random walks and self-supervised contrastive learning to develop a cyclical contrastive random walks (CCRW) method that distinguished salient anatomical regions from T2-weighted MRI, reducing human annotation for image segmentation. 

\subsection{Supervised Contrastive Learning for Multimodal Data}
In contrast to self-supervised contrastive learning approaches, supervised contrastive learning extends the conventional contrastive learning approaches to the fully-supervised setting by using data class label information \cite{khosla2020supervised}. Supervised contrastive learning, including Siamese network \cite{chopra2005learning}, Triplet network \cite{hoffer2015deep}, N-pair \cite{sohn2016improved}, SupCon \cite{khosla2020supervised, li2022learning, li2023generalized}, has achieved remarkable success. They have been applied to a number of medical imaging tasks \cite{zhang2021supervised,zhu2022deep,zhu2022multimodal}. In multimodal learning, supervised contrastive learning incorporated shared multimodal information from each subject to mine discriminative features for classification [ref]. For example, Ktena et al. \cite{ktena2017distance} proposed a Siamese graph convolutional network model to learn the similarity metric between irregular brain connectivities from heterogeneous rs-fMRI for autism diagnosis. Rossi et al. \cite{rossi2020multi} proposed a multimodal Siamese convolutional neural network to maximize the similarities of T2-weighted MRI and diffusion-weighted imaging data for prostate cancer diagnosis. Memmesheimer et al. \cite{memmesheimer2021sl} introduced a signal-level multimodal deep learning model using a Triplet network to project different skeleton sequences into a common feature space and then fed the learned fused features to a k-nearest neighbor model for action recognition. Zhang et al. \cite{zhang2021supervised} proposed supervised multimodal contrastive learning by applying a SupCon loss and a cross-entropy loss to jointly align image-text representation pairs for detecting unreliable news related to the Covid-19 pandemic. More recently, Zhu et al. \cite{zhu2022deep} took advantage of shared self-expression coefficients and generalized canonical correlation analysis to propose a multimodal discriminative and interpretability network for predicting Alzheimer’s disease using MRI, PET, and cerebrospinal fluid. Zhu et al. \cite{zhu2022multimodal} utilized a Triplet attention network to learn high-order discriminative features from rs-fMRI and DTI data to predict epilepsy disease.

\section{Methods}
\subsection{Overview}
\textbf{Figure 1} depicts the overview of our proposed multimodal feature integration method for early prediction of neurological deficits in very preterm infants. Suppose we have a training dataset $\bm{S}=\{\bm{s}^{(i)},y^{(i)}\}_{i=1}^{N}$ with $N$ subjects. We included three modalities of brain MRI data, including T2-weighted sMRI, DTI, fMRI, and clinical data. At the beginning, $m$ subjects are randomly sampled from the training dataset, i.e., $\Phi=\{1,2,\dots,m\}$ and $\bm{S}_{\Phi} \in \bm{S}$. For each subject $\bm{s}^{(i)}\in \bm{S}_{\Phi}$, let $x^{(i)}_{t}, x^{(i)}_{c}\in \bm{s}^{(i)}$ denotes the T2-weighted images and clinical data of a specific subject, respectively, we apply MRI preprocessing pipelines to parcellate the whole brain images into $d$ region of interests (ROIs), from which we extracted agnostic radiomic features $x^{(i)}_{r}\in \mathbb R^{d\times z}$, $z$ is the dimension of radiomic features of each ROI, and constructed brain structural connectome $x^{(i)}_{sc}\in \mathbb R^{d\times d}$ and functional connectome $x^{(i)}_{fc}\in \mathbb R^{d\times d}$, respectively. Note $\bm{s}^{(i)}=\{x^{(i)}_{r}, x^{(i)}_{sc}, x^{(i)}_{fc}, x^{(i)}_{t}, x^{(i)}_{c}\}$. After preprocessing, we obtained five different features/inputs. Next, a set of feature extractors $F(\cdot; \theta)$ is employed to map  $\bm{s}^{(i)}$ to $\textbf{f}^{(i)}, i.e.,  \textbf{f}^{(i)}=\{\text{f}^{(i)}_{r}, \text{f}^{(i)}_{sc}, \text{f}^{(i)}_{fc}, \text{f}^{(i)}_{t}, \text{f}^{(i)}_{c}\}$. Next, we design two pretext contrastive learning tasks to extract feature embeddings from five feature modalities to learn the \emph{CMC features} and the \emph{CSS features}. These two pretext tasks largely increase the training samples for the deep learning models, mitigating the inadequate data issue for model training in medical applications. Finally, we fine-tuned the pre-trained network to solve the downstream task (i.e., risk stratification of neurological deficits) in a supervised manner. Below we will elaborate on feature extraction, two pretext contrastive learning tasks, and other details. 
\begin{figure*}
    \centering
    \includegraphics[width=1.00\textwidth]{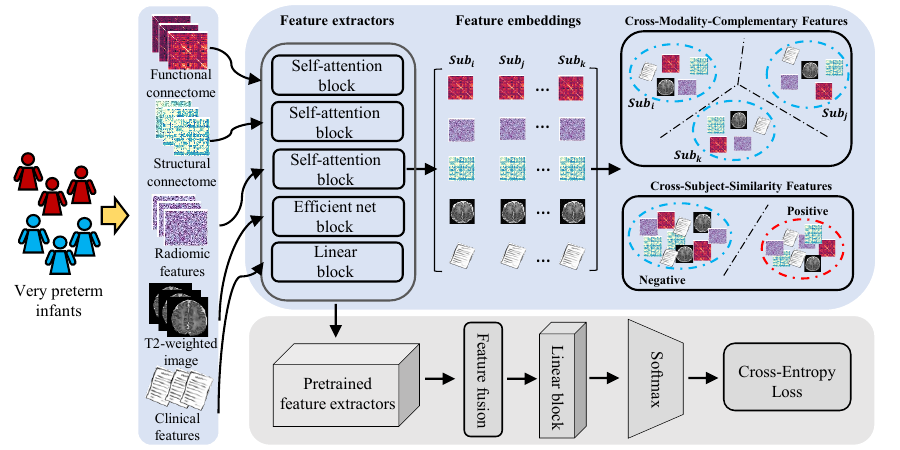}
    \caption{Schematic diagram of the proposed deep multimodal contrastive network for early prediction of neurological deficits at 2 years corrected age. We first input 5 feature types from $N$ subjects into a feature extractor block to extract the 5 different feature embeddings. Next, we performed two contrastive learning tasks to enforce the model to learn the CMC features and CSS features. Finally, we fine-tuned the pre-trained network in a supervised learning manner to predict the risk of cognitive deficits.}
    \label{framework}
\end{figure*}
\subsection{Feature Extraction}
Our proposed model was designed to equip five feature extractors to take five feature types from each subject. In brain imaging-based diagnosis, brain connectivity, e.g., structural connectome and functional connectome describe their unique characteristics, which can be utilized to analyze the spatial or sequential structure using the self-attention mechanism \citep{zhu2022multimodal}. For radiomic features, the self-attention mechanism can also be used to mine the implicit pathological information among different ROIs \citep{peng2017multilevel}. Let $\{W^{(i)}_q, W^{(i)}_k, W^{(i)}_v\}$ denote three parameter matrices for generating $i_{\text{th}}$ query $Q^{(i)}$, key $K^{(i)}$, and value $V^{(i)}$, respectively. Then, $\{Q^{(i)}, K^{(i)},V^{(i)}\}$ can be defined by different transformations on $\{x^{(i)}_{r}, x^{(i)}_{sc}, x^{(i)}_{fc}\}$ using linear mapping, which are
\begin{align}
    Q^{(i)}_u, Q^{(i)}_{u}, Q^{(i)}_{u} &= x^{(i)}_{u}W^{(i)}_{q}, x^{(i)}_{u}W^{(i)}_{q}, x^{(i)}_{u}W^{(i)}_{u}, u\in \{r,sc,fc\}
\end{align}

We then capture the attention score among different ROIs by computing the probability of scaled dot-product between $Q$ and $K$. Finally, the feature map with self-attention is calculated as another dot-product between the attention score and $V$, which is defined as follows:
\begin{align}
    A^{(i)}_{u} = \text{Softmax}(\frac{Q^{(i)}_u(K^{(i)}_u)^T}{\sqrt{d}})V^{(i)}_u, u\in \{r,sc,fc\}
\end{align}
where $A^{(i)}_{r}\in \mathbb R^{d\times p}, A^{(i)}_{sc} \in \mathbb R^{d\times d}, A^{(i)}_{fc} \in \mathbb R^{d\times d} $ are the self-attention map for $\{x^{(i)}_{r}, x^{(i)}_{sc}, x^{(i)}_{fc}\}$, respectively, and $d$ is a scaled parameter that equals to the number of ROIs. We employed a pre-trained EfficientNet \citep{tan2019efficientnet} and a fully connected network to extract image embedding from T2-weighted images and clinical embedding from clinical data. Finally, all attention maps, image embedding, and clinical embedding are followed by the same fully connected layer and a $L_2$ normalization layer, i.e., $\|\text{f}^{(i)}\|_2=1$, to obtain the high-level feature embeddings $\{\text{f}^{(i)}_{r}, \text{f}^{(i)}_{sc}, \text{f}^{(i)}_{fc}, \text{f}^{(i)}_{t}, \text{f}^{(i)}_{c}\}$, respectively.

\subsection{Learning Cross-Modality-Complementary Features}
\begin{figure}[ht]
    \centering
    \includegraphics[width=8.50cm]{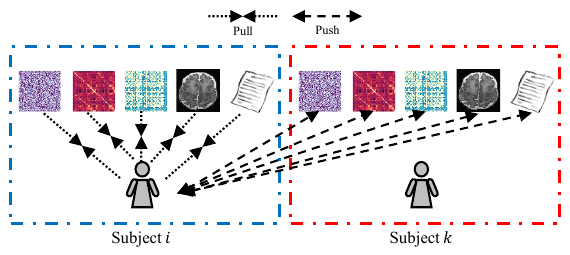}
    \caption{The illustration of learning CMC features from the proposed method.}
    \label{crossmodal}
\end{figure}

To reduce redundancy information and improve the complementary information among different modalities, we present a self-supervised contrastive learning pretext task to learn the \emph{CMC features} by mapping heterogeneous features into a common space for each subject (\textbf{Figure \ref{crossmodal}}). To achieve this, we randomly sample $m$ subjects, in which each subject consists of five feature types. Let $\{(x^{(1)}_{r}, \dots, x^{(1)}_{c}), \dots, (x^{(m)}_{r}, \dots, x^{(m)}_{c})\}$ denotes selected multimodal samples from $m$ subjects. These samples are fed into their corresponding feature extractors to get the high-level feature embeddings, i.e., $\{(\text{f}^{(1)}_{r}, \dots, \text{f}^{(1)}_{c}), \dots, (\text{f}^{(m)}_{r}, \dots, \text{f}^{(m)}_{c})\}$. Thus, the probability of $\bm{s}
^{(i)}=\{x^{(i)}_{r}, x^{(i)}_{sc}, x^{(i)}_{fc}, x^{(i)}_{t}, x^{(i)}_{c}\}$ being recognized as $i_{\text{th}}$ subject is defined by
\begin{align}
    p(i|\bm{s}^{(i)})=\frac{\sum_{u,v\in \{r,\dots,c\}}\exp\left[\text{f}^{(i)}_{u}(\text{f}^{(i)}_{v})^T/\tau\right]_{(u\neq v)}}{\sum_{j\in \Phi}\sum_{u,v\in \{r,\dots,c\}}\exp\left[\text{f}^{(i)}_{u}(\text{f}^{(j)}_{v})^T/\tau\right]_{(u\neq v,i\neq j)}}
\end{align}
where $\text{f}^{(i)}_{u}(\text{f}^{(i)}_{v})^T$ denotes the cosine similarity between $\text{f}^{(i)}_{u}$ and $\text{f}^{(i)}_{v}$, indicating two modalities are arise from a specific subject. $\tau$ denotes a temperature parameter, which controls the density level of sample distribution. In experiments, we empirically set $\tau$ to 1 \cite{chen2020simple,li2021rotation}.

Meanwhile, the distance between each feature embedding of a subject should be mutually exclusive. Therefore, similar to \textbf{Eq (3)}, the probability of $\bm{s}
^{(k)}=\{x^{(k)}_{r}, x^{(k)}_{sc}, x^{(k)}_{fc}, x^{(k)}_{t}, x^{(k)}_{c}\}$ being recognized as $i_{\text{th}}$ subject is defined by
\begin{align}
    p(i|\bm{s}^{(k)})=\frac{\sum_{u,v\in \{r,\dots,c\}}\exp\left[\text{f}^{(i)}_{u}(\text{f}^{(k)}_{v})^T/\tau\right]_{(u\neq v,i\neq k)}}{\sum_{j\in \Phi}\sum_{u,v\in \{r,\dots,c\}}\exp\left[\text{f}^{(j)}_{u}(\text{f}^{(k)}_{v})^T/\tau\right]_{(u\neq v, j\neq k)}}
\end{align}
Now assume that all probabilities of different samples being recognized as $i_{\text{th}}$ subject are independent, the objective likelihood function, such that $\bm{s}^{(i)}$ being recognized as $i_{\text{th}}$ subject and $\bm{s}^{(k)}$ not being recognized as $i_{\text{th}}$ subject is defined as
\begin{align}
     \ell_{cmc}=\prod_{i\in \Phi}\prod_{k\in \Phi} p(i|\bm{s}^{(i)})\left[1-p(i|\bm{s}^{(k)})\right]
\end{align}
Thus, the \emph{CMC loss} $\mathcal{L}_{cmc}$ is defined as the negative-log-likelihood of $\ell_{cmc}, i.e, \mathcal{L}_{cmc}=-\log \ell_{cmc}$, which can be simplified to
\begin{align}
    \mathcal{L}_{cmc} = -\frac{1}{|\Phi|}\left(\sum_{i\in \Phi}\log p(i|\bm{s}^{(i)}) - \sum_{i\in \Phi}\sum_{k\in \Phi}\log p(i|\bm{s}^{(k)})\right)
\end{align}
where $|\Phi|=m$ denotes the size of $\Phi$. Thus, we learn the \emph{CMC features} by grouping the feature embeddings of an individual subject and separating each subject from other subjects.

\subsection{Learning Cross-Subject-Similarity Features}
\begin{figure}[ht]
    \centering
    \includegraphics[width=12.00cm]{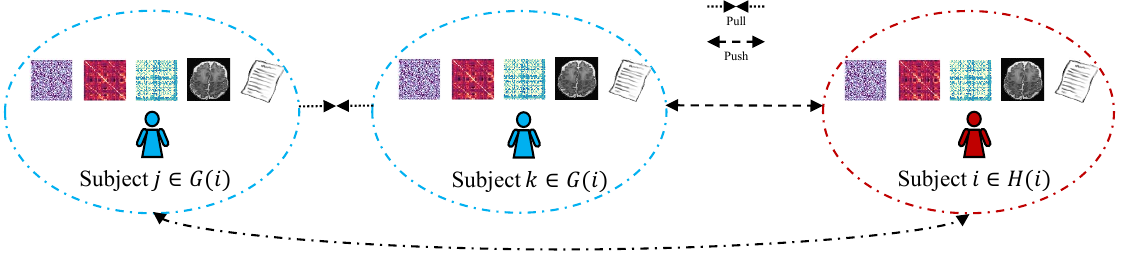}
    \caption{The illustration of learning CSS features from the proposed method.}
    \label{crosssubject}
\end{figure}

The cross-subject data modalities should share similar information if their corresponding subjects have the same disease outcomes. This concept is shown in \textbf{Figure \ref{crosssubject}}. We learn \emph{CSS features} to improve the alignment of similar subjects for instance discrimination. Let $G(i)=\{j\in \Phi|y^{(i)}=y^{(j)}, i\neq j\}, G(i)\in \bm{S}_{\phi}$ denote the set of indices for the samples with the same label.
The probability of $\bm{s}^{(i)}$ and $\bm{s}^{(g)}$ sharing a same disease outcome, e.g., $y^{(i)}=y^{(g)}, i\neq g$ is defined as 
\begin{align}
    p(y^{(i)}=y^{(g)}|\bm{s}^{(i)},\bm{s}^{(g)})=\frac{\sum_{u,v\in \{r,\dots,c\}}\exp\left[\text{f}^{(i)}_{u}(\text{f}^{(g)}_{v})^T/\tau\right]_{(u\neq v)}}{\sum_{j\in \Phi}\sum_{u,v\in \{r,\dots,c\}}\exp\left[\text{f}^{(i)}_{u}(\text{f}^{(j)}_{v})^T/\tau\right]_{(u\neq v,i\neq j)}}
\end{align}
Then, the \emph{CSS} $\mathcal{L}_{css}$ can be expressed as follows:
\begin{align}
    \mathcal{L}_{css}=-\frac{1}{|\Phi|}\sum_{i\in \Phi}\frac{1}{|G(i)|}\sum_{g\in G(i)}\log p(y^{(i)}=y^{(g)}|\bm{s}^{(i)},\bm{s}^{(g)}) 
\end{align}
Minimizing $\mathcal{L}_{css}$ yields the purpose of learning \emph{CSS features} through a supervised contrastive learning approach that separates subjects with different disease outcomes and groups them with the same disease outcomes.

\subsection{Representation Joint Learning Objective}
Our learning objective is defined as a weighted linear combination of two contrastive loss functions to learn both \textbf{CMC features} and \textbf{CSS features}. Thus, the learning objective loss is formulated as follows:
\begin{align}
    \mathcal{L}^{*}=\lambda \mathcal{L}_{cmc} + \mathcal{L}_{css}
\end{align}
where $\lambda$ is the weighting factor for controlling the relative importance of $\mathcal{L}_{cmc}$, respectively. In experiments, similar to \cite{li2021rotation,dai2021adaptive}, $\lambda=1$ shows the best classification performance. We also investigate the effects of different $\lambda$ in the ablation study section. For our real downstream task, we fused each pre-trained feature extractor and fine-tuned the fused embeddings with a fully-connected layer. We then employed a Softmax function and a weighted cross-entropy loss to perform a downstream classification task.  In general, the overview of our proposed method is summarized in \textbf{Algorithm 1}.
\begin{algorithm}
  \small
  \caption{Proposed Method}
  \begin{algorithmic}[1]
    \Inputs{$\bm{S}_{\Phi}=\{(x^{(i)}_{r}, x^{(i)}_{sc}, x^{(i)}_{fc},x^{(i)}_{t}, x^{(i)}_{c}),y^{(i)}\}_{i=1}^{N}$}
    \Initialize{Shared weights $W^{(i)}_q, W^{(i)}_k, W^{(i)}_v$}
    \While{epoch $<$ MaxEpochs}
        \For{$m\in N, i\in \Phi$}
            \For{$u\in \{r,sc,fc,t,c\}$}
                \State $Q^{(i)}_u, K^{(i)}_{u}, V^{(i)}_{u} \gets x^{(i)}_{u}W^{(i)}_{q}, x^{(i)}_{u}W^{(i)}_{k}, x^{(i)}_{u}W^{(i)}_{v}$
                \State $A^{(i)}_{u}\gets \text{Softmax}(\frac{Q^{(i)}_u(K^{(i)}_u)^T}{\sqrt{d}})V^{(i)}_u$
                \State $\text{f}^{(i)}_{u}\gets \text{Norm}\left(\text{MLPs}(A^{(i)}_{u})\right)$
                \If{u=t}
                    \State $A^{(i)}_{u}\gets \text{EfficientNet}(x^{(i)}_t)$
                    \State $\text{f}^{(i)}_{u}\gets \text{Norm}\left(\text{MLPs}(A^{(i)}_{u})\right)$
                \EndIf
                \If{u=c}
                    \State $\text{f}^{(i)}_{u}\gets \text{Norm}\left(\text{MLPs}(x^{(i)}_t)\right)$
                \EndIf
                \State Save each $\text{f}^{(i)}_{u}$
            \EndFor
            \State Compute $\mathcal{L}_{cmc}\left(\text{f}^{(i)}_{u},\text{f}^{i}_{v}\right)$ by \textbf{Eq.(6)}
            \State Compute $\mathcal{L}_{css}\left(\text{f}^{(i)}_{u},\text{f}^{(i)}_{v},y^{(i)}\right)$ by \textbf{Eq.(8)}
            \State $\mathcal{L}^{*}=\lambda \mathcal{L}_{cmc} + \mathcal{L}_{css}$
            \State Update network
        \EndFor
    \EndWhile
    \State \Return Each pre-trained feature extractor
  \end{algorithmic}
\end{algorithm}

\subsection{Network Implementation Details}
As shown in \textbf{Figure \ref{framework}}. For all subjects in a batch (b=32), we applied three self-attention networks, with the same architecture as the \cite{vaswani2017attention}, to extract attention maps (size of 87 x 87) from functional connectome, structural connectome, and radiomic features, respectively. We applied a fully connected layer with 10 nodes to these attention maps to reduce the dimensions of all attention maps to 87x10. We then flatted these attention maps and applied two fully-connected layers with nodes of 256 and 128, respectively, to extract the feature embeddings. For T2-weighted MRI images, we selected 10 slices of whole brain T2-weighted image volume and resized them to 224 x 224. We then used a pre-trained 3D EfficientNet backbone \cite{tan2019efficientnet} and applied the last fully-connected layer with 128 nodes. For clinical data, we used a fully-connected layer with 128 nodes to extract the features from the perinatal clinical information. To obtain the same-sized feature embeddings from all feature extractors, we used another fully-connected layer to map each feature embedding to the same size of 8. After that, we jointly trained our feature extractors with two contrastive learning loss functions. Finally, we added a fusion layer to fuse all feature representations from individual feature extractors, and a fully-connected layer (2 nodes) with Softmax function as the model output. We fine-tuned the whole model for the downstream classification task using a weighted cross-entropy loss in a supervised manner. Same as \cite{ li2021rotation}, the network is optimized using the Adam optimizer with a learning rate of 0.001 and a weight decay of 0.001. We train our two contrastive learning tasks and downstream tasks for 2000 and 500 epochs, respectively. The whole framework was implemented using Python 3.8, Scikit-Learn 0.24.1, Pytorch 1.9.1, and Cuda 11.1 with a NVIDIA GeForce GTX 1660 SUPER GPU.

\section{Data and Experimental Results}
\subsection{CINEPS Dataset}
We developed and validated our model using a regional prospective cohort of very preterm infants from the Cincinnati Infant Neurodevelopment Early Prediction Study (CINEPS) \cite{parikh2021perinatal}. Subjects with known congenital brain anomalies or severe perinatal injury were excluded, resulting in 300 labeled subjects from the CINEPS cohort. For MRI acquisition, all subjects were imaged at 39-44 weeks postmenstrual age during unsedated sleep on the same 3T Philips Ingenia scanner using a 32-channel receiver head coil at Cincinnati Children’s Hospital Medical Center (CCHMC). sMRI data were scanned using a T2-weighted turbo spin-echo protocol. rs-fMRI data were collected using a multi-brand (factor=3). DTI data were collected using single-shot planar imaging. Detailed MRI scanning acquisition parameters can be found in prior literature \cite{kline2022diffuse,kelly2023neuroimaging}. As the gold standard reference of neurodevelopmental deficits, each subject was assessed at 2 years corrected age using the Bayley Scales of Infant and Toddler Development, 3rd Ed. (Bayley-III) test \cite{bayley2006bayley} with neurodevelopmental scores ranging from 40 to 145 in the cohort. We dichotomized subjects into two groups: the low-risk group (score$\ge$85, N=192) and the high-risk group (score$<$85, N=108) for cognitive scores, the low-risk group (score$\ge$85, N=75) and the high-risk group (score$<$85, N=222) for motor scores, and the low-risk group (score$\ge$85, N=94) and the high-risk group (score$<$85, N=202) for motor scores.

\subsection{COEPS Dataset}
Our model is validated using an external independent dataset via the Columbus Early Prediction Study (COEPS), which includes 83 subjects from Nationwide Children’s Hospital (NCH). We excluded the subjects with congenital or chromosomal anomalies that impact the central nervous system. Subjects were scanned at 38–43 weeks PMA on the same 3T MRI scanner (Skyra; Siemens Health- care) with a 32-channel pediatric head coil at NCH. Detailed MRI scanning acquisition parameters can be found in prior literature \cite{li2022novel,li2023novel}. Bayley III tests were also conducted to collect the cognitive score for all subjects at 2 years of corrected age. Similar to the CINEPS dataset, we dichotomized subjects into two groups and obtained 68 subjects in the low-risk group (cognitive score$\ge$85) and 15 subjects in the high-risk group (cognitive score$<$85).

\subsection{MRI Data Preprocessing and Postprocessing}
The original T2-weighted images were processed using the developing Human Connectome Project (dHCP) pipeline \cite{makropoulos2018developing} to segment whole brain images into 87 regions of interest (ROIs) based on an age-matched neonatal volumetric atlas \cite{gousias2012magnetic}. The full description of 87 ROIs can be found in their original paper. In general, the dHCP pipeline first applies developing brain region annotation with expectation maximization (Draw-EM) algorithm \cite{gousias2012magnetic,makropoulos2018developing} to segmented T2-weighted MRI images into 9 tissues (e.g., cortical grey matter and white matter), and then performed a multi-channel registration approach to register labeled neonatal atlases with 87 ROIs to each subject. For each ROI, we extracted a total of 100 agnostic radiomic features using the PyRadiomics pipeline \cite{van2017computational}, resulting in a 2D radiomic feature map for each subject. We preprocessed DTI and rs-fMRI data using the corresponding dHCP pipelines \cite{makropoulos2018developing}. We constructed brain structural connectome by treating 87 ROIs of age-matched neonatal atlas as graph nodes and FA-weighted fiber tract counts as graph edges. Meanwhile, we constructed a functional connectome by considering those 87 ROIs as graph nodes and correlation among ROIs’ BOLD signals as graph edges. Additional details can be found in prior literature \cite{gousias2012magnetic,makropoulos2018developing}. Together with T2-weighted original images and perinatal clinical data collected prior to neonatal intensive care unit discharge, we obtained five different types of features in total for model input.

\subsection{Experimental Setting}
\subsubsection{Competing Multimodal Learning Approaches}
We compared the proposed method with peer conventional deep multimodal fusion, self-supervised contrastive learning, and supervised contrastive learning approaches. To have a fair comparison, we trained all competing methods on the same feature extractors, batch size, optimizer, learning rate, and weight decay term. 

1) \textbf{Deep-Multimodal} \cite{he2021deep}. 
We previously proposed a deep multimodal learning model to predict the neurological deficits of very preterm infants. To apply this model in our study, we concatenated the extracted feature embeddings and added a fully-connected layer to reduce the fused features’ dimensions to 2 for classification. The model was trained using the cross-entropy loss. We treated the Deep-Multimodal method as the baseline method in our study.

2) \textbf{Weighted-DCCA} \cite{liu2019multimodal}. 
The weighted DCCA method applies the CCA constraint to regulate the non-linear mappings of extracted features from different modalities. In this study, we applied the same feature extractors as the proposed method and specified the CCA constraint to maximize the correlation between multimodal features. Since we have five inputs and CCA is originally proposed for two variable input sets, we accordingly set the CCA constraint for each pair combination. Next, same as \cite{liu2019multimodal}, we fused each extracted feature using weighted summation with a convex linear combination, following a fully-connected layer with 2 nodes for classification. The model was trained using a cross-entropy loss.

3) \textbf{Deep sr-DDL} \cite{d2021deep}.
The Deep sr-DDL method was proposed to predict the clinical outcomes using dynamic correlation matrices. In this study, we retained the input features and feature extractors, but only made changes in the last fully connected layer by reducing the number of nodes to 2. In addition, we replaced the MSE loss with cross-entropy loss. 

4) \textbf{Modality-Invariant} \cite{li2020self}.
The Modality-Invariant method utilizes self-supervised learning techniques to capture semantically shared information among synthesized modalities. To compare with our multimodal method, we kept the feature extractors the same as ours and applied the Modality-Invariant method in our input feature to pre-train the feature extractors. Finally, we used the same approach as our method in fine-tuning the whole modeling stage for classification after the pretraining network using the Modality-Invariant method. 

5) \textbf{MRI-Siamese} \cite{rossi2020multi}.
The core idea of the MRI-Siamese method is to capture the pairwise similarity between representations of two subjects with the same class label from network encoders. In our study, we first fused the extracted features from all input types, and we further applied the MRI-Siamese method to learn the discriminative features by maximizing the agreement for a pair subject with the same class label. After that, we fine-tuned the pre-trained feature extractors using the same approach as the proposed method for classification. 

6) \textbf{MRI-Triplet} \cite{zhu2022multimodal}.
The MRI-Triplet was proposed based on a triplet network for brain disease diagnosis using multiple MRI data. In this study, we adopted a triplet network on the fused feature embeddings for classification. We pre-trained the network with a joint loss function of triplet loss and cross-entropy loss. The same as our proposed method, we further fine-tuned the pre-trained network for classification. 

\subsubsection{Model Evaluation Strategy}
We evaluated the proposed and other competing methods using binary classification metrics. In particular, balanced accuracy (BA), sensitivity (SEN), specificity (SPE), and the area under the receiver operating characteristic (ROC) curve (AUC) were applied to evaluate classification performance. As an internal validation using the CINEPS dataset, we conducted a 10-fold cross-validation. In each iteration, we set 9 subsets of the entire dataset as training data, and the remaining subset was treated as independent testing data. Training data (i.e., 9 subsets) were further split into training data for model training and validation data for model optimization. The model with the best validation loss was selected across all training epochs and tested on unseen testing data. We repeated this process for 10 iterations until each subset of the cohort was used as testing data. We then repeated this cross-validation process 50 times and reported the mean metrics and their standard deviation (SD) to evaluate performance variances. To show the generalizability of our method, we tested an internally validated model from the CINEPS dataset using the unseen independent COEPS dataset. A non-parametric Wilcoxon test was applied with a p-value less than 0.05 for all statistical inferences to show the statistical significance of completing methods. We conducted all statistical tests in R-4.0.3 (RStudio, Boston, MA, USA).

\subsection{Internal Validation on CINEPS Dataset}
\begin{table}[ht]
    \centering
    \caption{The internal valuation of early prediction of cognitive deficits using different competing methods on CINEPS dataset (Experimental results are represented as mean $\pm$ SD).}
    \begin{tabular}{c|c|c|c|c}
    \hline
     	& BA (\%)  & AUC (\%) & SEN (\%) & SPE (\%)\\
     \hline
     Deep-Multimodal &  66.8 $\pm$ 3.0 & 65.3 $\pm$ 4.2 & 64.3 $\pm$ 4.6 & 69.2 $\pm$ 3.4\\
     Weighted-DCCA    &  68.3 $\pm$ 4.3 & 69.5 $\pm$ 4.9 & 67.2 $\pm$ 5.0 & 71.7 $\pm$ 4.5\\
     Deep sr-DDL     &  65.0 $\pm$ 3.2 & 63.5 $\pm$ 3.7 & 61.2 $\pm$ 4.2 & 68.5 $\pm$ 3.8 \\
     Modality-Invariant & 77.3 $\pm$ 3.9 &	78.4 $\pm$ 5.1 & 76.3 $\pm$ 4.5 & 78.2 $\pm$ 4.0\\
     MRI-Siamese     & 75.1 $\pm$ 4.6 &	74.6 $\pm$ 6.8 & 73.5 $\pm$ 5.4 & 76.7 $\pm$ 4.2 \\
     MRI-Triplet     & 77.4 $\pm$ 3.7 &	77.0 $\pm$ 4.5 & 75.7 $\pm$ 4.6 & 79.0 $\pm$ 3.9\\
     \textbf{Ours} & \textbf{82.4 $\pm$ 4.6} & \textbf{81.5 $\pm$ 5.6} & \textbf{80.5 $\pm$ 5.4} & \textbf{84.3 $\pm$ 4.5}\\
     \hline
    \end{tabular}
    \label{Tab1}
\end{table}
\begin{table}[ht]
    \centering
    \caption{The internal valuation of early prediction of motor deficits using different competing methods on CINEPS dataset (Experimental results are represented as mean $\pm$ SD).}
    \begin{tabular}{c|c|c|c|c}
    \hline
     	& BA (\%)  & AUC (\%) & SEN (\%) & SPE (\%)\\
     \hline
     Deep-Multimodal &  68.9 $\pm$ 4.7 & 65.5 $\pm$ 5.1 & 67.3 $\pm$ 4.3 & 67.8 $\pm$ 4.2\\
     Weighted-DCCA    &  68.7 $\pm$ 5.2 & 67.2 $\pm$ 4.6 & 66.8 $\pm$ 5.3 & 70.5 $\pm$ 4.7\\
     Deep sr-DDL     &  66.4 $\pm$ 4.3 & 63.5 $\pm$ 3.7 & 63.4 $\pm$ 4.7 & 69.3 $\pm$ 4.2 \\
     Modality-Invariant & 76.1 $\pm$ 4.1 &	73.2 $\pm$ 5.1 & 72.7 $\pm$ 4.5 & 79.4 $\pm$ 3.8\\
     MRI-Siamese     & 73.1 $\pm$ 4.5 &	71.4 $\pm$ 5.6 & 70.3 $\pm$ 4.9 & 75.8 $\pm$ 3.9 \\
     MRI-Triplet     & 75.2 $\pm$ 4.9 &	72.8 $\pm$ 4.8 & 72.9 $\pm$ 5.2 & 77.4 $\pm$ 4.6\\
     \textbf{Ours} & \textbf{78.3 $\pm$ 4.6} & \textbf{76.1 $\pm$ 6.1} & \textbf{75.2 $\pm$ 5.7} & \textbf{81.3 $\pm$ 5.2}\\
     \hline
    \end{tabular}
    \label{Tab2}
\end{table}
\begin{table}[ht]
    \centering
    \caption{The internal valuation of early prediction of language deficits using different competing methods on CINEPS dataset (Experimental results are represented as mean $\pm$ SD).}
    \begin{tabular}{c|c|c|c|c}
    \hline
     	& BA (\%)  & AUC (\%) & SEN (\%) & SPE (\%)\\
     \hline
     Deep-Multimodal &  67.9 $\pm$ 4.9 & 64.7 $\pm$ 5.5 & 66.8 $\pm$ 4.9 & 68.9 $\pm$ 4.4\\
     Weighted-DCCA    &  68.1 $\pm$ 4.2 & 65.4 $\pm$ 5.2 & 67.3 $\pm$ 5.3 & 69.0 $\pm$ 3.9\\
     Deep sr-DDL     &  63.6 $\pm$ 3.5 & 62.0 $\pm$ 3.7 & 61.8 $\pm$ 4.1 & 65.4 $\pm$ 3.5 \\
     Modality-Invariant & 73.6 $\pm$ 3.8 & 73.2 $\pm$ 5.1 & 69.7 $\pm$ 4.3 & 77.4 $\pm$ 4.1\\
     MRI-Siamese     & 71.9 $\pm$ 4.0 &	69.8 $\pm$ 5.5 & 67.5 $\pm$ 4.5 & 76.2 $\pm$ 3.7 \\
     MRI-Triplet     & 72.8 $\pm$ 4.5 &	70.0 $\pm$ 5.2 & 68.0 $\pm$ 4.9 & 77.5 $\pm$ 4.3\\
     \textbf{Ours} & \textbf{75.6 $\pm$ 4.9} & \textbf{73.4 $\pm$ 5.3} & \textbf{72.0 $\pm$ 6.5} & \textbf{79.1 $\pm$ 5.7}\\
     \hline
    \end{tabular}
    \label{Tab3}
\end{table}
\begin{figure*}
    \centering
    \includegraphics[width=1.00\textwidth]{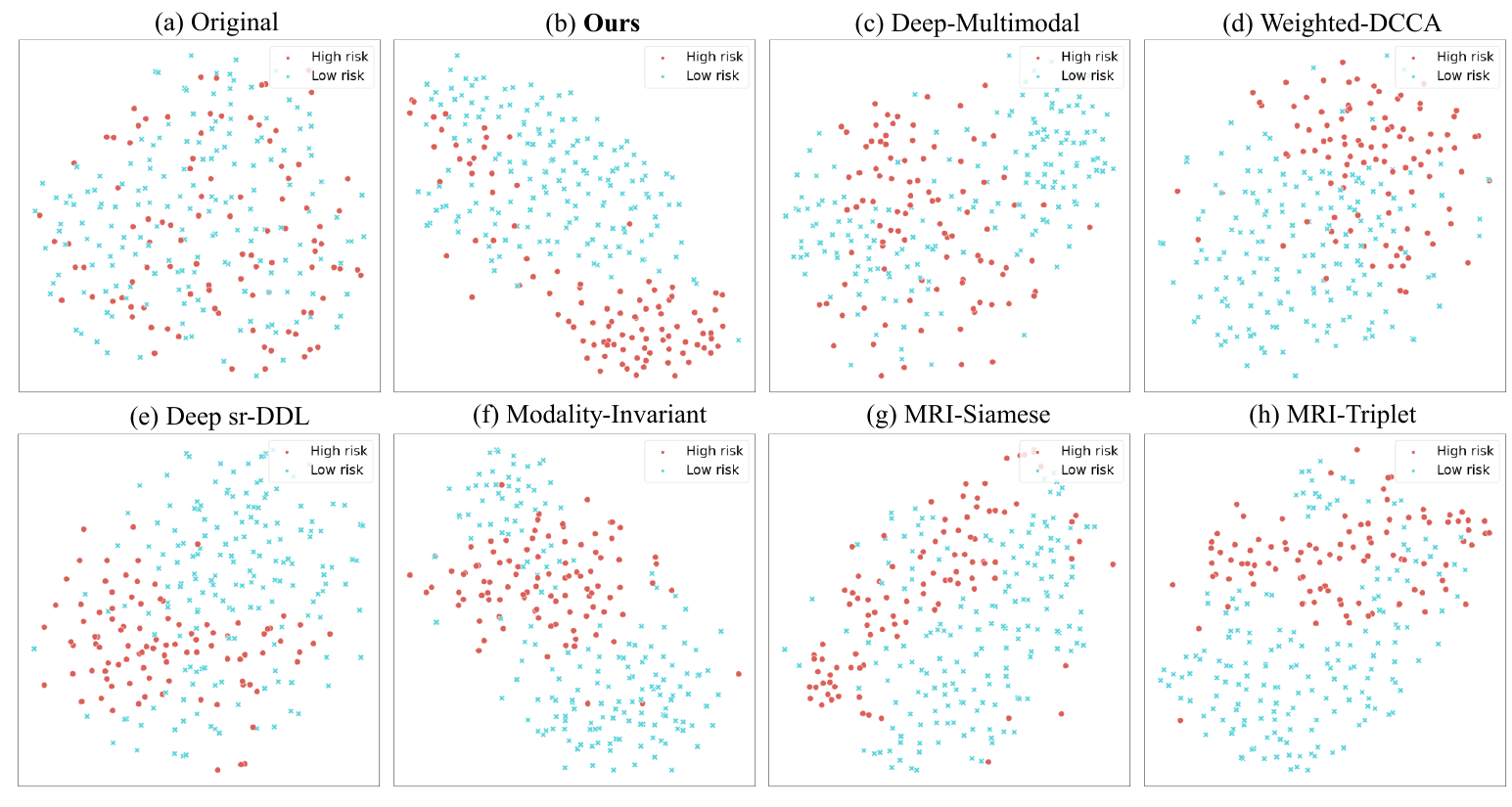}
    \caption{The t-SNE visualization of different methods for prediction of cognitive deficits uses the network’s last hidden layer in latent feature space. (a) is the feature representation in the original space before model optimization (b) is the feature representation learned from our method, we used the last hidden layer in the downstream stage. (c-h) are feature representations learned from other competing methods.}
    \label{tsne1}
\end{figure*}
The risk stratification of cognitive deficits of our proposed method and other competing methods are shown in \textbf{Table \ref{Tab1}}. Our method achieved the best classification results in the prediction of cognitive deficits with 82.4\% on BA, 81.5\% on AUC, 80.5\% on SEN, and 84.3\% on SPE. Likewise, our methods also outperform other methods in risk stratification of motor and language deficits (\textbf{Table \ref{Tab2}-\ref{Tab3}}). These experimental results indicate that our method learns better representative features than other competing methods. We plotted the learned feature representation using the t-SNE plot in \textbf{Figure \ref{tsne1}}. Visually, it is easier to separate the latent feature representation of our methods with a clearer decision boundary than other competing methods. Compared to the second-best method Modality-Invariant \cite{li2020self}, our method significantly improved the performance of cognitive deficits diagnosis by around 3.1\% (p$<$0.001) on AUC and 6.1\% (p$<$0.001) on BA. In addition, our method significantly outperforms the baseline method Deep-Multimodal \cite{he2021deep} by 16.2\% on AUC and 16.6\% on BA. These results further demonstrated the effectiveness of our method.

\subsection{External Validation on COEPS Dataset}
\begin{figure*}
    \centering
    \includegraphics[width=0.65\textwidth]{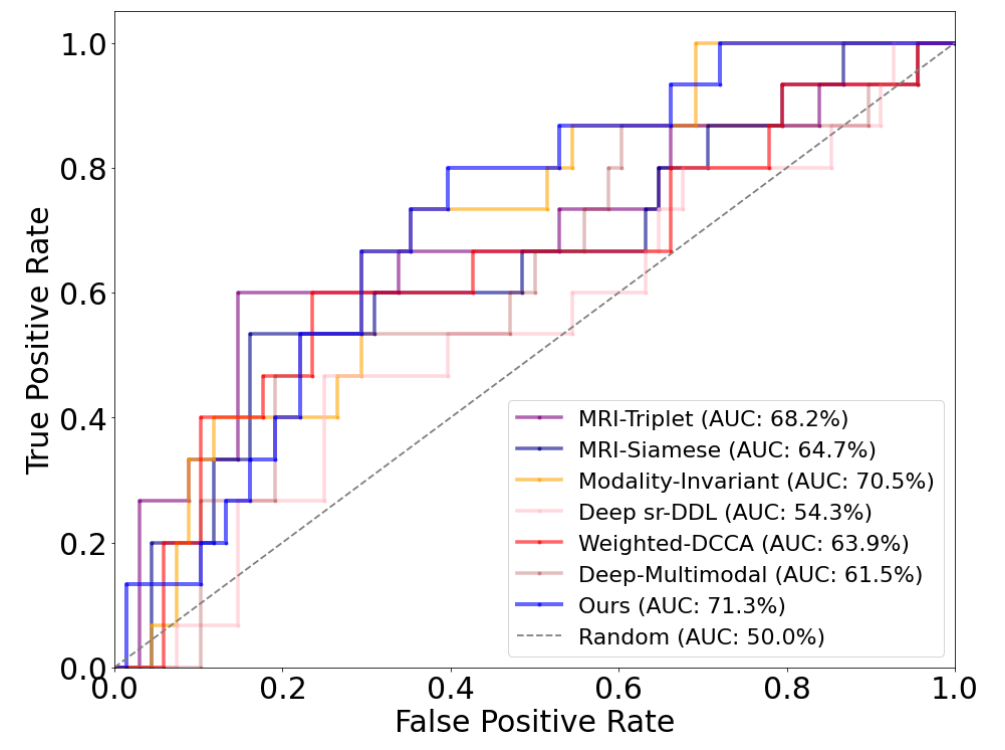}
    \caption{The ROC curves of different competing methods. The AUC values are shown in the lower right of the figure.}
    \label{ROC}
\end{figure*}

To show the generalizability of our method, we trained each method on the CINEPS dataset and employed an independent COEPS dataset to externally validate each model. The results are shown in  \textbf{Table \ref{Tab4}}. Similar to internal validation, Modality-Invariant achieved the second-best results with 67.9\% on BA and 70.5\% on AUC. Our method surpassed other competing methods with 68.6\% on BA and 71.3\% on AUC. We also provided the ROC curves of individual methods in the ROC curves (\textbf{Figure \ref{ROC}}). These external results provided the generalization capability of our method.

\begin{table}[ht]
    \centering
    \caption{The external valuation of early prediction of cognitive deficits using different competing methods on COEPS dataset.}
    \begin{tabular}{c|c|c|c|c}
     \hline
     	& BA (\%)  & AUC (\%) & SEN (\%) & SPE (\%)\\
     \hline
     Deep-Multimodal & 60.5 & 61.5 & 53.3 &	67.6 \\
     Weighted-DCCA   & 63.8 & 63.9 &	60.0  &	67.6 \\
     Deep sr-DDL    & 54.2 & 	54.3 & 	46.7 & 	61.8 \\
     Modality-Invariant & 67.9 & 70.5 &	\textbf{66.7} &	69.1 \\
     MRI-Siamese & 63.4 &	64.7 &	53.3 &	\textbf{73.5} \\
     MRI-Triplet   & 65.3 &	68.2 &	60.0 & 70.6 \\
     \textbf{Ours} & \textbf{68.6} & \textbf{71.3} & \textbf{66.7} & 70.6 \\
     \hline
    \end{tabular}
    \label{Tab4}
\end{table}

\subsection{Ablation Study}
\subsubsection{Effects of Cross-Modality-Complementary Features}
Our method combines CMC loss $\mathcal{L}_{cmc}$ and CSS loss $\mathcal{L}_{css}$. $\mathcal{L}_{cmc}$ is used to align each feature into a common space to learn the complementary information from different modalities. Depending on the effectiveness of alignment, the classification performance of the downstream task may vary. Therefore, we analyzed the importance of learning the CMC features by training our method with different $\lambda$ in \textbf{Eq (9)}. $\lambda = 0.00$ indicates that the method excludes $\mathcal{L}_{cmc}$, achieves an AUC of 75.5\% and a BA of 76.5\%. As $\lambda$ increases, the model starts to obtain a better classification performance until $\lambda$ reaches 1.00. When $\lambda$ keeps increasing, the classification performance starts to drop down to 75.0\% on AUC and 76.2\% on BA. We can see that our method achieved the best classification results with $\lambda = 1.00$, demonstrating the equal contribution of $\mathcal{L}_{cmc}$ and $\mathcal{L}_{css}$.
\begin{table}[ht]
    \centering
    \caption{The effects of the CMC loss $\mathcal{L}_{cmc}$. $\lambda$ indicates a weighting factor of $\mathcal{L}_{cmc}$ in in \textbf{Eq (9)}. We analyzed the classification results based on different $\lambda$ on the CINEPS dataset.}
    \begin{tabular}{c|c|c|c|c}
     \hline
      & BA (\%) & AUC (\%) & SEN (\%) & SPE (\%) \\
      \hline
      $\lambda=0.00$ & 76.5 $\pm$ 3.8 & 75.5 $\pm$ 4.6 & 74.8 $\pm$ 4.2 & 78.2 $\pm$ 4.0 \\
      $\lambda=0.50$ & 76.8 $\pm$ 4.3 & 77.5 $\pm$ 4.8 & 76.1 $\pm$ 4.9 & 77.4 $\pm$ 4.5 \\
      $\lambda=0.75$ & 80.0 $\pm$ 4.6 & 78.0 $\pm$ 4.6 & 79.5 $\pm$ 5.1 & 80.5 $\pm$ 4.5 \\
      $\lambda=1.00$ & \textbf{82.4 $\pm$ 4.6} & \textbf{81.5 $\pm$ 5.6} & \textbf{80.5 $\pm$ 5.4} & \textbf{84.3 $\pm$ 4.5}\\
      $\lambda=1.50$ & 78.0 $\pm$ 4.0 & 77.0 $\pm$ 4.2 & 77.5 $\pm$ 4.8 & 78.4 $\pm$ 4.3 \\
      $\lambda=2.00$ & 76.2 $\pm$ 4.5 & 75.0 $\pm$ 4.8 & 75.3 $\pm$ 5.2 & 77.0 $\pm$ 4.6 \\
     \hline
    \end{tabular}
    \label{Tab5}
\end{table}

\subsubsection{Effects of Individual Loss}
To analyze the effects of each individual contrastive loss, we compared the classification performance for identifying cognitive deficits using the CINEPS dataset. The results are shown in \textbf{Table \ref{Tab6}}. We can see that the model trained only with a CMC loss $\mathcal{L}_{cmc}$ loss excelled over the model trained only with a CSS loss $\mathcal{L}_{css}$, i.e., 79.3\% vs 76.5\% on BA and 78.2\% vs 75.5\% on AUC. This observation also supports the results of \textbf{Table \ref{Tab3}} that effectively project heterogeneous features into a common space helps the model learn the complementary information from different modalities and further enhances the classification performance. The model that was jointly trained with two contrastive loss functions achieved the best classification performance, demonstrating the effectiveness of capturing the synergistic effect created by modalities and subjects.

\begin{table}[ht]
    \centering
    \caption{Performance comparison for cognitive deficits risk stratification on the CINEPS dataset using CMC loss alone, CSS loss alone, and combined loss.}
    \begin{tabular}{ccccc}
     \hline
      & BA (\%) & AUC (\%) & SEN (\%) & SPE (\%) \\
      \hline
      Cross-Modality & 79.3 $\pm$ 4.6 & 78.2 $\pm$ 5.1 & 78.0 $\pm$ 5.2 & 80.5 $\pm$ 4.8 \\
      Cross-Subject & 76.5 $\pm$ 3.8 & 75.5 $\pm$ 4.6 & 74.8 $\pm$ 4.2 & 78.2 $\pm$ 4.0 \\
      \textbf{Ours} & \textbf{82.4 $\pm$ 4.6} & \textbf{81.5 $\pm$ 5.6} & \textbf{80.5 $\pm$ 5.4} & \textbf{84.3 $\pm$ 4.5}\\
     \hline
    \end{tabular}
    \label{Tab6}
\end{table}
\begin{figure*}[ht]
    \centering
    \includegraphics[width=1.00\textwidth]{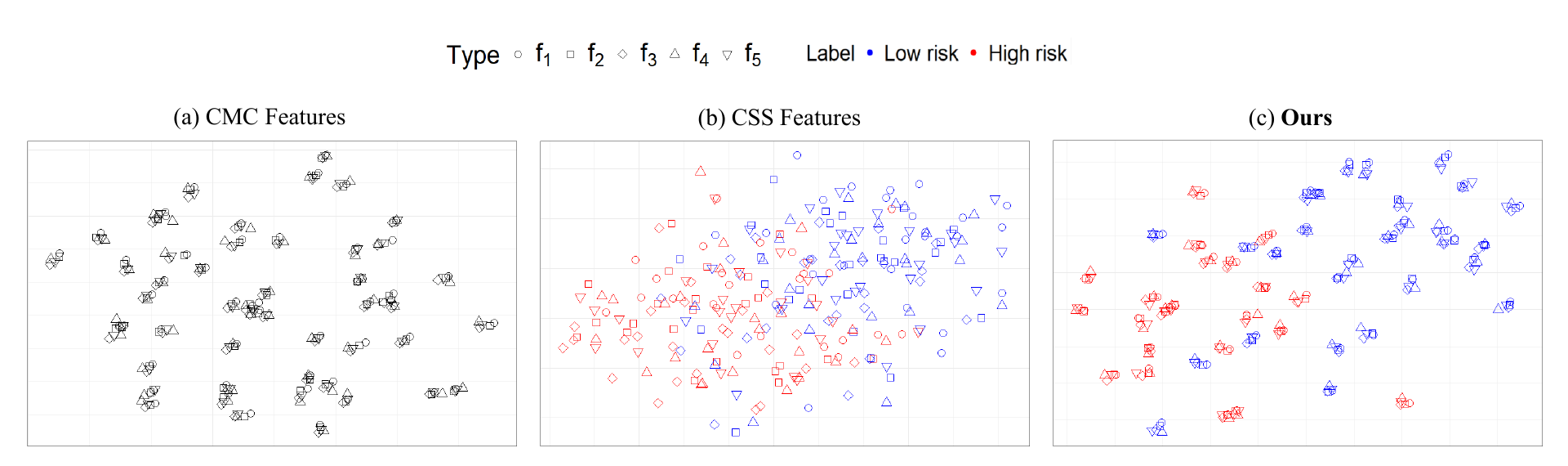}
    \caption{The t-SNE visualization of heterogeneous features from different modalities. We visualized each t-SNE based on different contrastive loss functions. The closer heterogeneous features are embedded, the better complementary information is captured. The closer similar features are embedded, the better discriminative features are learned. ($\text{f}_1,\text{f}_2,\text{f}_3,\text{f}_4,\text{f}_5$ represent the feature representation from functional connectome, structure connectome, radiomics, T2-weighted images, and clinical features, respectively.)}
    \label{tsne2}
\end{figure*}
\subsubsection{Feature Visualization}
To compare the effectiveness of learned latent feature representation using different contrastive loss, we used a t-SNE plot to visualize the latent features from different modalities on the CINEPS dataset. As shown in \textbf{Figure \ref{tsne2}}, optimizing the CMC loss successfully maps heterogeneous features into a common space (\textbf{Figure \ref{tsne2}a}). This learning process captures complementary information and reduces noise redundancy across modalities. \textbf{Figure \ref{tsne2}b} shows that by optimizing the CSS loss, features of subjects with the same class labels were pulled together, and features with different were pushed away. The model with joint loss functions not only captured complementary information and reduced the noise redundancy across modalities, but also learned the discriminative features by considering the similarities across subjects (\textbf{Figure \ref{tsne2}c}). These results further support the classification results using the CINEPS dataset in \textbf{Table \ref{Tab1}}, demonstrating the superiority of the learned feature representations of our method.

\begin{figure*}[ht]
    \centering
    \includegraphics[width=1.00\textwidth]{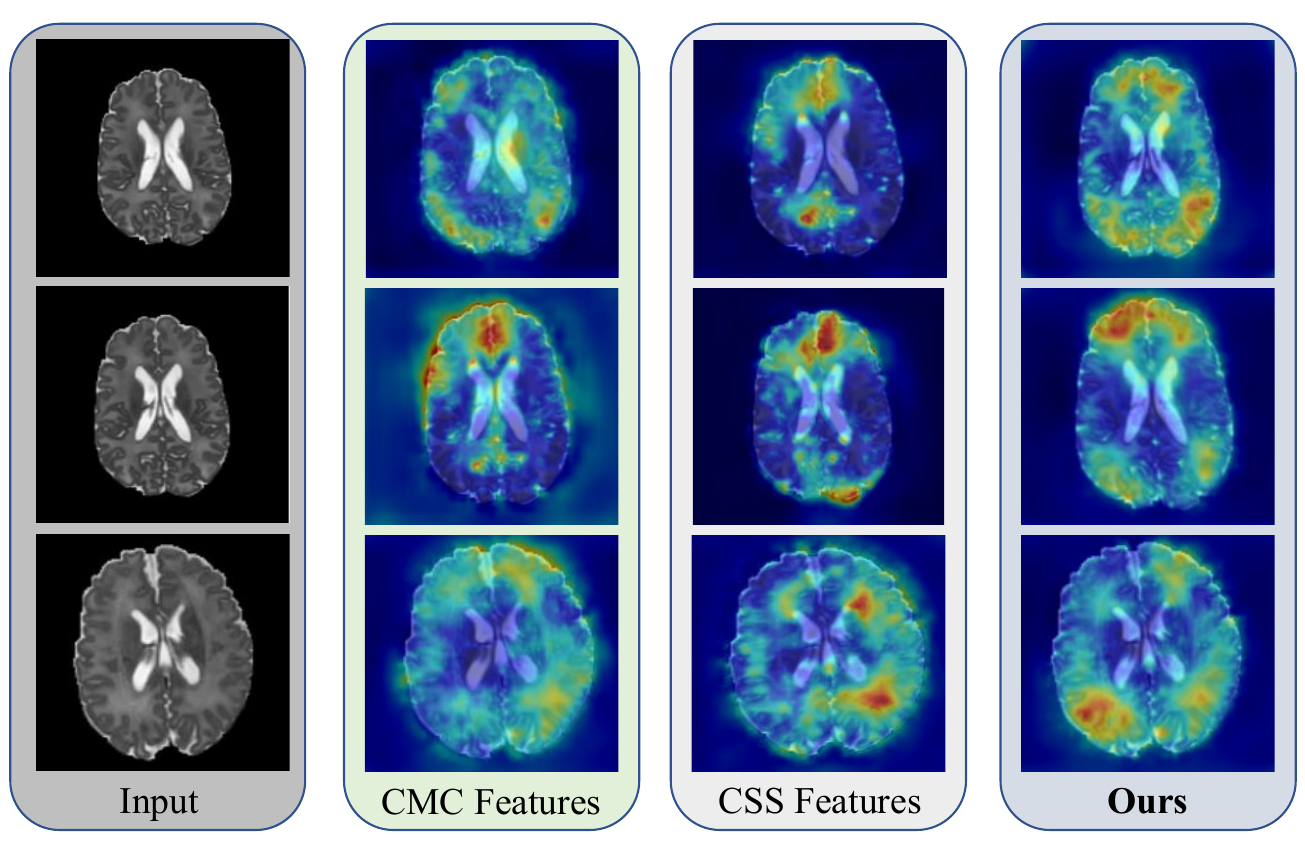}
    \caption{Grad-CAM visualization of three slice examples in T2-weighted images from the proposed method. We compared our method with each individual contrastive loss function (e.g., CMC loss and CSS loss). Each heatmap highlights the discriminative regions using red color, which corresponds to high-value scores in the Grad-CAM heatmap.}
    \label{cam}
\end{figure*}

In addition, we visualized learned imaging features on T2-weighted images to verify whether the proposed method is able to successfully recognize the anatomy patterns that are related to cognitive deficits diagnosis. We explained our models using the Grad-CAM \cite{selvaraju2017grad} to visualize the heatmap from the EfficientNet block. The results are illustrated in \textbf{Figure \ref{cam}}. The Grad-CAM heatmap of the proposed method shows regions more complete regions, while training CMC loss and CSS loss separately only localizes partial discriminative regions. This visualization further demonstrates that our method can effectively learn the discriminative features of predicting cognitive deficits using T2-weighted images. 

\begin{landscape}
 \begin{table}
    \caption{Impact of individual modality (fMRI to Clinical data) on classification performance for cognitive deficits using the CINEPS dataset. We removed one input feature and used the rest of the four feature types in the proposed method.}
    \begin{tabular}{c|c|c|c|c|c|c|c|c|c}
    \hline
    Trial & fMRI & DTI & Radiomics & T2 & Clinical & BA (\%) & AUC (\%) & SEN (\%) & SPE (\%) \\
    \hline
    1 & & \checkmark  & \checkmark  & \checkmark  & \checkmark  & 79.6 $\pm$ 5.8 & 77.3 $\pm$ 6.7 & 76.8 $\pm$ 6.2 & 82.4 $\pm$ 5.7 \\
    2 & \checkmark &   & \checkmark  & \checkmark  & \checkmark  & 79.2 $\pm$ 4.8 & 75.0 $\pm$ 7.5 & 75.2 $\pm$ 6.5 & 80.1 $\pm$ 5.5 \\
    3 & \checkmark &   \checkmark &  & \checkmark  & \checkmark  & 74.8 $\pm$ 4.4 & 72.0 $\pm$ 6.5 & 73.5 $\pm$ 5.9 & 76.1 $\pm$ 5.4 \\
    4 & \checkmark &   \checkmark &   \checkmark &  & \checkmark  & 81.0 $\pm$ 5.1 & 80.0 $\pm$ 5.5 & 79.5 $\pm$ 6.7 & 82.4 $\pm$ 5.5 \\
    5 & \checkmark &   \checkmark & \checkmark & \checkmark  &  & 78.0 $\pm$ 4.4 & 76.7 $\pm$ 4.8 & 76.5 $\pm$ 6.2 & 79.5 $\pm$ 4.8 \\
    \hline
    \textbf{All} & \checkmark &  \checkmark  & \checkmark  & \checkmark  & \checkmark & \textbf{82.4 $\pm$ 4.6} & \textbf{81.5 $\pm$ 5.6} & \textbf{80.5 $\pm$ 5.4} & \textbf{84.3 $\pm$ 4.5}\\
    \end{tabular}
    \label{Tab7}
 \end{table}
\end{landscape}
\subsubsection{Impact of Individual Modality}
In this section, we provided an ablation study to evaluate the impact of individual modality on the prediction of cognitive deficits (\textbf{Table \ref{Tab7}}). Particularly, we excluded one input modality and used the rest of the four modalities. For example, in the first trial, we excluded the fMRI modality and retained all other input features (sMRI, radiomics, etc.). We observed that the proposed method achieved the highest classification results with a balanced accuracy of 81.0\% and AUC of 80.0\% in trial 4, in which we excluded all T2-weighted images. This indicates that the T2-weighted MRI image modality has the least impact among all five input features. On the other hand, the model excluding radiomic features achieved the lowest classification performance in trial 3, demonstrating that radiomic features have the largest impact on predicting cognitive deficit such a phenomenon is also observed in another ablation study, where we explored the prediction power using each individual modality. (\textbf{Table \ref{Tab8}}) We trained CSS loss on each individual feature to learn the CSS features on the CINEPS dataset. As shown in \textbf{Table \ref{Tab8}}, the model using radiomic features achieved the highest classification performance with a balanced accuracy of 75.0\% and an AUC of 75.3\%. Our method including all modalities achieved the best classification results, showing that each modality has its own contribution to discovering the discriminative features in the prediction of cognitive deficits.

\begin{table}[ht]
    \centering
    \caption{Performance comparison using individual modality to predict cognitive deficits.}
    \begin{tabular}{ccccc}
     \hline
      Modality & BA (\%) & AUC (\%) & SEN (\%) & SPE (\%) \\
      \hline
     T2 & 69.7 $\pm$ 4.5 & 67.5 $\pm$ 5.2 & 68.1 $\pm$ 6.5 & 71.2 $\pm$ 5.3	\\
     Clinical & 72.5 $\pm$ 4.3 & 69.6 $\pm$ 4.9 & 71.5 $\pm$ 6.2 & 73.5 $\pm$ 5.1	\\
     DTI & 74.3 $\pm$ 4.5 & 73.4 $\pm$ 6.1 & 72.0 $\pm$ 6.3 & 76.5 $\pm$ 5.2 \\
     fMRI & 74.0 $\pm$ 4.3 & 71.5 $\pm$ 6.5 & 72.9 $\pm$ 6.0 & 75.0 $\pm$ 5.5 \\
     Radiomics & 75.0 $\pm$ 4.8 & 75.3 $\pm$ 7.1 & 74.3 $\pm$ 6.5 & 75.5 $\pm$ 5.8 \\
     \textbf{Ours} & \textbf{82.4 $\pm$ 4.6} & \textbf{81.5 $\pm$ 5.6} & \textbf{80.5 $\pm$ 5.4} & \textbf{84.3 $\pm$ 4.5}\\
     \hline
    \end{tabular}
    \label{Tab8}
\end{table}

\section{Discussion}
Early prediction of neurological deficits in very preterm infants continues to be a challenging task in clinical practice. An accurate prognostic classifier is desired to facilitate risk stratification and prevent the absence of prompt treatment for children. In the neuroimaging study, multimodal MRI data, such as sMRI, DTI, and fMRI, provides complementary information about unique characteristics of the brain, which further improves the accuracy of neurodevelopment abnormalities diagnosis \cite{dai2020multimodal,lee2022multimodal}. With the advances in deep learning techniques, multimodal learning with multiple MRI data has been studied to explore to enhance the prediction performance of neurodevelopmental impairments in very preterm neonates \cite{he2021deep} by integrating relevant brain features from different MRI modalities. However, conventional multimodal learning methods naively fuse these heterogeneous feature representations that are located in different representation spaces, resulting in complementary information not being appropriately captured \cite{bakkali2023vlcdoc}. Self-supervised contrastive learning approaches, including CLIP-based methods \cite{radford2021learning,sanghi2022clip,wang2022clip},  successfully capture complementary information by projecting multimodal feature representation into a common space, where the heterogeneous features can be effectively combined. Meanwhile, supervised contrastive learning techniques, such as the Siamese network \cite{bromley1993signature}, Triplet network \cite{hoffer2015deep}, and SupCon \cite{khosla2020supervised}, incorporate shared information among different subjects by pulling similar subjects and pushing away dissimilar subjects to reduce the redundancy of multimodal data. 

In this work, we proposed a novel joint self-supervised and supervised contrastive learning method to amalgamate complementary information across modalities via CMC features and shared information across subjects via CSS features for early prediction of neurological deficits in very preterm infants. Learning CMC features helps our model enhance the complementary semantics and reduce the redundancy from different modalities. Meanwhile, learning CSS features helps our model identify the commonalities between different subjects and mine the discriminative features for classification. Our method has been validated on two independent datasets, i.e., CINEPS and COEPS datasets, for early prediction of neurodevelopmental abnormalities. Our method consistently achieved the best prediction performance among other competing multimodal learning, self-supervised, and supervised contrastive learning methods. There are some other methods, such as Deep-Multimodal \cite{he2021deep}, Weighted-DCCA \cite{liu2019multimodal}, and Deep sr-DDL \cite{d2021deep}, which were proposed for learning multimodal data but achieved limited performance in this study. This is due to the fact that these methods do not map heterogeneous features into a common space, resulting in ineffective fusion of multimodal features and reducing their redundancy. Modality-Invariant \cite{li2020self} considers fusing multimodal features to capture the commentary information by mapping them into a common space but ignores the shared information across subjects. On the other hand, MRI-Siamese \cite{rossi2020multi} and MRI-Triplet \cite{zhu2022multimodal} incorporate shared information across subjects to enhance the classification performance but the complementary information among different modalities was disregarded.

We analyzed our method in various ablation studies. We considered the importance of learning CMC and CSS features and provided an analysis of different weighting factors $\lambda$ in \textbf{Eq (9)}. The results from \textbf{Table \ref{Tab3}} show that the proposed method achieved the highest prediction performance when $\lambda=1.00$, indicating the equal contribution of $\mathcal{L}_{cmc}$ and $\mathcal{L}_{css}$. Such phenomenon was also shown in \textbf{Table \ref{Tab4}} that jointly training $\mathcal{L}_{cmc}$ and $\mathcal{L}_{css}$ helps the model to capture the synergistic effect created by both modalities and subjects. To interpret why the proposed methods can have superior performance for neurodevelopmental abnormalities diagnosis, we showed feature visualization of learned latent feature representations of the proposed method using t-SNE plots in \textbf{Figure \ref{tsne2}}. We observed that our method captures successfully embedded multimodal feature representations together and learns better discriminative features. In addition, we applied Grad-CAM to visualize the learned imaging features in \textbf{Figure \ref{cam}}, in which our method precisely captures the discriminative regions for making decisions to diagnose neurodevelopmental abnormalities in very preterm neonates.  Furthermore, we analyzed the impact of individual modalities on the prediction of neurological deficits (\textbf{Table \ref{Tab5} \& \ref{Tab6}}). We obtained that the imaging modality has the least impact among other modalities while radiomics has the largest impact on the prediction of neurological deficits in very preterm infants. 

Our work contains some limitations. First, we only considered the scenario that all modalities of a subject are available in the current study. In reality, some modalities might be missing or only contain a few samples. In the future, we will investigate how to apply the multimodal fusion method to address the missing modalities problem. Second, our model is evaluated on the CINEPS dataset that contains 300 labeled subjects. This can be considered a large dataset in the neuroimaging study, but still limited for deep learning models. In the future, we will consider using additional unlabeled data to address the small-sized labeled data problem. Finally, our external validation only contains 83 subjects on the COEPS dataset, of which 15 subjects were from the high-risk group. This could affect the classification performance since the label is imbalanced. Moving forward, we will also need to evaluate our method with a large external dataset for robustness and generalizability purposes.

\section{Conclusion}
In this paper, we proposed a novel joint self-supervised and supervised contrastive learning method on multimodal MRI data for early prediction of neurological deficits in very preterm infants. Our main idea is to effectively capture complementary information and reduce redundancy to enhance the synergistic effect of different modalities and subjects by learning the \emph{cross-modality-complementary} features and \emph{cross-subject-similarity} features. Our method was validated on extensive experiments, demonstrating the effectiveness of our learned fused features for neurological deficit diagnosis. With further refinement, the proposed method may facilitate computer-aided diagnosis in using multimodal data in clinical practice.

\section*{Acknowledgement}
This work was supported by the National Institutes of Health [R01-EB029944, R01-EB030582, R01-NS094200, and R01-NS096037]; Academic and Research Committee (ARC) Awards of Cincinnati Children’s Hospital Medical Center.

\section*{Ethics Statement}
In accordance with The Code of Ethics of the World Medical Association, this study was approved by the Institutional Review Boards of the Cincinnati Children's Hospital Medical Center (CCHMC) and Nationwide Children's Hospital (NCH). Written parental informed consent was obtained for each subject.

\bibliographystyle{elsarticle-num}
\bibliography{ref}
\end{document}